# Strong Visible Absorption and Photoluminescence of Titanic Acid Nanotubes by Hydrothermal Method

B. L. Tian,[†,‡] X. T. Zhang,[†] S. X. Dai,[†] K. Cheng,[†,§] Z. S. Jin,[†] Y. B. Huang,[†] Z. L. Du,[*,†] G. T. Zou,[‡] and B. S. Zou[§]

*Key Lab for Special Functional Materials of Ministry of Education, Henan University, Kaifeng 475004, P. R. China, National Lab of Superhard Materials, Jilin University, Changchun, 130012, P. R. China, and Nanophysics and Nanodevice Lab, Institute of Physics, Chinese Academy of Sciences, Beijing 100080, P. R. China*

Titanic acid nanotubes (with a chemical formula $H_2Ti_2O_4(OH)_2$, abbreviated as TANTs) were synthesized by the hydrothermal method using commercial $TiO_2$ nanoparticle powder (P25, Degussa, Germany) including anatase and rutile phase as a starting material. Conversion from nanoparticles to nanotubes was achieved by treating the nanoparticle powder with 10 M NaOH aqueous solution. Absorption and photoluminescence (PL) data indicate that the nanotubes obtained under slow and suitable drying and heating conditions had very strong and stable visible absorption with three peaks at 515, 575, and 675 nm and photoluminescence at room temperature in air.

**Introduction**

Titanium dioxide as a kind of large band gap semiconductor material has been widely studied in recent years due to its potential applications in photochemical water splitting,[1,2] solar energy conversion systems,[3,4] various electro-optical devices,[5] photocatalytic systems,[6] photonic crystals,[7] degradation of environmental contaminants,[8] etc. However, the lack of absorption in the visible region prevented applications on utilizing sunlight. More efforts have been carried out in improving its visible absorption through dye-sensitizing[9−11] and -doping[12,13] methods. By those methods, the optic−electronic response of $TiO_2$ can extend to the visible region. However, the stability and spectral responding character of those materials in air needs to further meliorate before any applications. It is well known that nanostructured materials have special properties than their bulk materials with a decrease of dimension. Since the discovery of carbon nanotubes by Iijima,[14] one-dimensional nanostructured materials with exceptional electrical, optical, and mechanical properties and potential applications have been extensively studied. Besides carbon nanotubes, many other one-dimensional nanostructured materials, such as III−V compound semiconductors nanowires,[15,16] CdTe nanowires,[17] ZnO nanobelts,[18] and C−BN−C coaxial nanocable,[19] have been successfully fabricated. In recent years, there has been growing interest in the preparation of $TiO_2$-based nanotube materials because titanate nanotubes can be synthesized readily.[20−25]

In 1998, Kasuga et al.[20] first obtained a kind of tubular material that was 8 nm in diameter and 100 nm in length by treating $80TiO_2 \cdot 20SiO_2$ (in mol %) powders with a 5−10 M NaOH aqueous solution. We also obtained tubular materials with an inner diameter of 4.2−5.9 nm and a length of about a few hundred nanometers from $TiO_2$ nanoparticles (P-25) using an analogous and improved method in 2000.[21] In former work these nanotubular materials were regarded as "$TiO_2$ nanotubes". However, Peng et al.[22] suggested that it was nanotube $TiO_x$ or $H_2Ti_3O_7$. After further investigation we found that the so-called "$TiO_2$ nanotubes" were actually titanate sodium nanotubes which could be converted to titanic acid nanotubes in HCl solution, and the chemical formula could be written as $Na_2Ti_2O_4(OH)_2$ and $H_2Ti_2O_4(OH)_2$, respectively.[23]

By far, tens of papers on the properties of the titanate nanotubes have been reported. However, there are few reports on the unusual optical properties of titanate nanotubes in the visible region. It is found that these tubular materials had bigger specific surface area than that of $TiO_2$ nanoparticles (P-25) and could be used as novel photoelectric materials.[21] Uchida et al.[24] and Adachi et al.[25] found that the nanotubes could be applied to a dye-sensitized solar cell system and had better sunlight−electricity conversion. Recently we found that the $H_2Ti_2O_4(OH)_2$ nanotubes show special optical properties in the visible region[26] and ESR response.[27] On further investigation, it was found that titanic acid nanotubes had strong and stable absorption and photoluminescence phenomenon in the visible region at room temperature in air. In this paper we present the results from this study.

**Experimental Section**

Starting materials were $TiO_2$ nanoparticles powder (P25, Degussa Co., Germany). It included anatase (80%) and rutile (20%) phase. The nanotubes were prepared through the hydrothermal treatment of $TiO_2$ nanoparticles powder in 10 M NaOH solution under stirring for 24 h at 120 °C. The obtained dispersion was cooled to room temperature and the upper liquid poured off. Then the residual dispersion was rinsed in turn with deionized water, acid solution, and deionized water, then dried at 60−70 °C for 48 h under vacuum, and finally annealed at 100, 200, 300, 400, and 500 °C for 2 h in air. The preparation process was the same as that of our previous works[26,27] expect for the drying and heating conditions.

---

* To whom correspondence should be addressed. E-mail: zld@henu.edu.cn.
† Henan University.
‡ Jilin University.
§ Institute of Physics, Chinese Academy of Sciences.

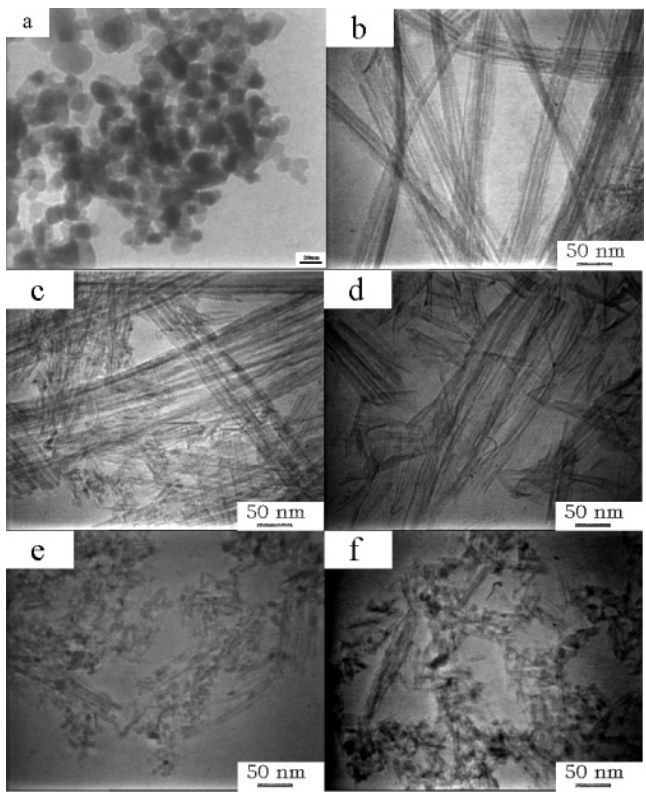

**Figure 1.** TEM images of raw material P-25 and the samples annealed at different temperatures for 2 h: (a) P-25; (b) 100 °C; (c) 200 °C; (d) 300 °C; (e) 400 °C; (f) 500 °C.

Examination of the samples' morphology was performed using a transmission electron microscope (TEM JEM-2010). The samples were dispersed by ultrasonication in ethanol and then placed on the perforated carbon copper grids. X-ray diffraction (XRD) patterns of the samples were measured by an X'Pert Pro X-ray diffractometer with Cu Kα ($\lambda = 1.54$ Å) radiation. The diffuse reflectance spectra (DRS) were measured with a TU-1901 UV−Visible spectrophotometer (reference substance, $BaSO_4$). PL spectra of all samples were measured using an ultraviolet−visible spectrophotometer (SPEX F212) with a Xe lamp as the excitation light source at room temperature in air. All specimens measured were powders.

### Results and Discussion

Figure 1 shows TEM photographs of the raw materials, and the samples were annealed at different temperature for 2 h. From Figure 1a we could observe that the grain diameter of the raw material is about 20−30 nm. It is also obvious that at the low heating temperatures (shown in Figure 1b−d) these tubular materials were commonly multiwalled, of which typical size was several nanometers in inner diameter and a few hundred nanometers in length. However, at higher heating temperatures (Figure 1e, f), we found the tubular structure was badly destroyed.

In order to better understand their structural behavior, XRD experiments were performed. Figure 2 shows XRD spectra of the raw materials and the samples annealed at different temperatures for 2 h. It is obviously seen that at low heating temperatures (Figure 2a, b, and c) the tubular materials had three diffraction peaks at 9.18°, 24.30°, and 48.14°. They can be indexed to (200), (110), and (020) crystal planes of the orthorhombic system (JCPDS, 47-0124), respectively. However, at 400 °C (Figure 2d) it is accompanied with formation of anatase phase and the nanotube morphology was destroyed. In addition, the XRD pattern of 500 °C (Figure 2e) annealed nanotubes is similar to that of the raw material (Figure 2f). The intensities of the diffraction peaks belonging to the anatase phase are increased. These results were in agreement with the above TEM results.

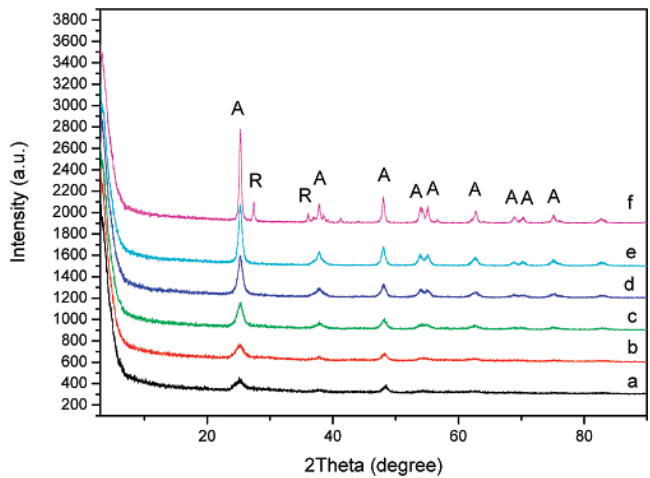

**Figure 2.** XRD spectra of the raw material P-25 and the samples annealed at different temperatures for 2 h: (a) 100 °C; (b) 200 °C; (c) 300 °C; (d) 400 °C; (e) 500 °C; (f) P-25 (A, anatase; R, rutile).

To our great surprise those tubular materials showed extremely special absorption characteristic. According to the Kubelka−Munk (KM) theory for the diffuse reflectance of a powder sample, the diffuse reflection spectra of the titanic acid nanotubes are transformed into the absorption proportional function. The KM function is as follows[28,29]

$$F(R_\infty) = (1 - R_\infty)^2/2R_\infty = K/S \quad (1)$$

where constants $K$ and $S$ characterize the losses of incident light due to absorption and scattering, respectively. Since $S$ is independent of wavelength, so $F(R_\infty)$ is proportional to the real absorption of sample

$$F(R_\infty) = (1 - R_\infty)^2/2R_\infty \propto \text{absorption} \quad (2)$$

The plot of $F(R_\infty)$ as a function of wavelength is shown in Figure 3. It is confirmed that the direct correlation between absorption and wavelength is reasonable. From Figure 3 we could observe the nanotubes had strong visible absorption compared to P-25 nanoparticles. Especially, in the region of 100−300 °C heating, the sample appeared to have three strong visible absorption peaks at 515, 575, and 675 nm (see Figure 3a, b, and c, respectively). When the heating temperature increases further, at 400 and 500 °C, the peaks disappeared and the absorption intensity in visible region decreased (see Figure 3d, e).

Notice that the tubular structure was already destroyed when heating temperature was above 400 °C observed by TEM and XRD. Obviously, the novel absorption in the visible region of the nanotubes was from its special tubular structure. On the other hand, we also noted that the relative absorption intensity of three peaks of the nanotubes varies irregularly with heating temperature, namely, at lower temperature. At 100 °C, the sample showed weaker absorption and the intensity of the three peaks (515, 575, and 675 nm) increased gradually (Figure 3a); at 200 °C, the sample showed strong absorption and the relative magnitude of the three peaks intensity was completely reversed (Figure 3b); while at 300 °C, the relative magnitude recovered

5363

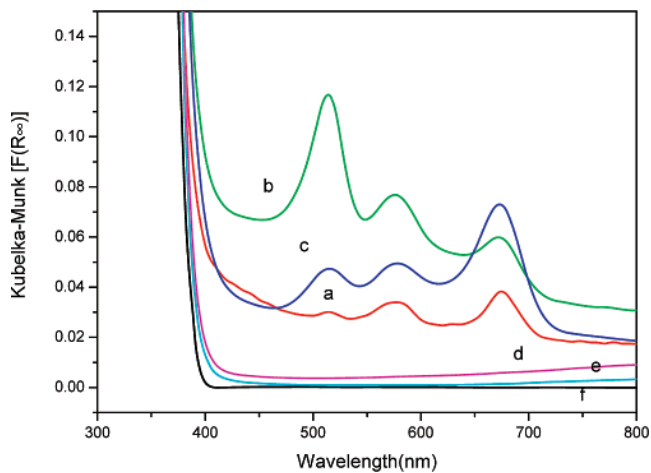

**Figure 3.** Wavelength dependence of $F(R_\infty)$ for P-25 and the samples annealed at different temperatures for 2 h: (a) 100 °C; (b) 200 °C; (c) 300 °C; (d) 400 °C; (e) 500 °C; (f) P-25.

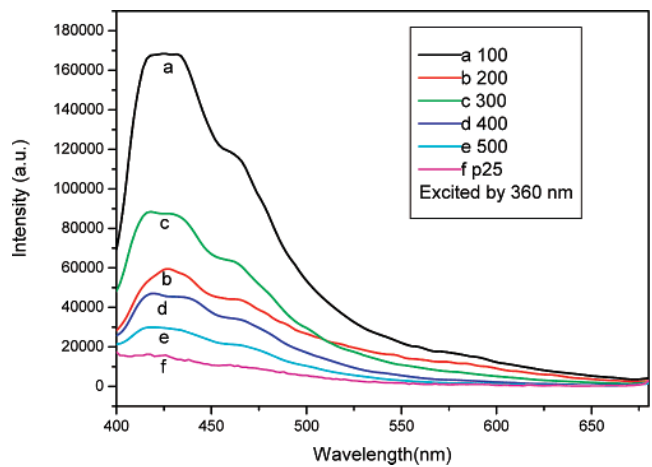

**Figure 5.** PL spectra excited by 360 nm of P-25 and the samples annealed at different temperatures for 2 h: (a) 100 °C; (b) 200 °C; (c) 300 °C; (d) 400 °C; (e) 500 °C; (f) P-25.

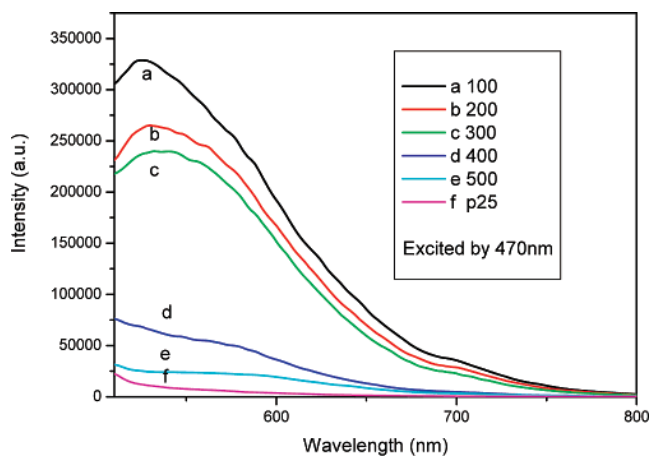

**Figure 4.** PL spectra excited by 470 nm of P-25 and the samples annealed at different temperatures for 2 h: (a) 100 °C; (b) 200 °C; (c) 300 °C; (d) 400 °C; (e) 500 °C; (f) P-25.

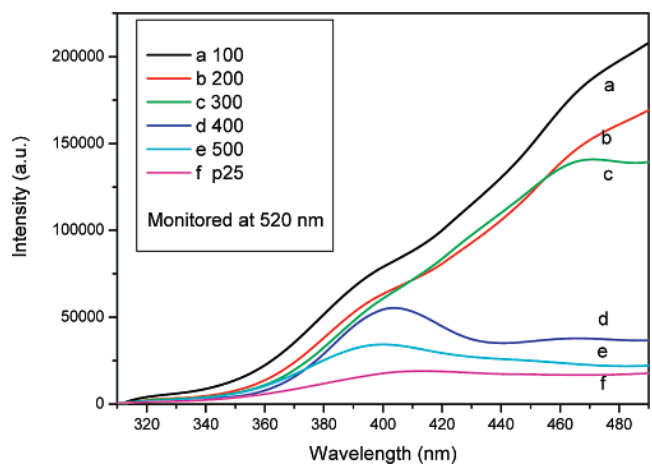

**Figure 6.** PLE spectra monitored at 520 nm of P-25 and the samples annealed at different temperatures for 2 h: (a) 100 °C; (b) 200 °C; (c) 300 °C; (d) 400 °C; (e) 500 °C; (f) P-25.

again (Figure 3c) and became similar to that heated at 100 °C. They implied that the optical property of the tubular materials was very sensitive to the heating temperature.

However, it is also found that TANTs had very strong photoluminescence excited by visible light. Figure 4 shows the PL spectra of TANTs heated at different temperature and P-25 in which the excitation wavelength was 470 nm. From the figure one can clearly observe that the samples heated at 100, 200, and 300 °C (Figure 4a, b, and c, respectively) had a very strong PL peak at about 540 nm with a weak shoulder at about 700 nm at the excitation of 470 nm radiations. However, for the samples heated at 400 and 500 °C (Figure 4d and e), the PL intensity was very weak and no peak at about 540 nm appeared. In fact, the PL spectra heated at 400 and 500 °C were rather similar to that of P-25 (Figure 4f). Moreover, it is found that the strong PL peak of the nanotubes could vary from 420 to 650 nm with changing of excitation wavelength, and this also suggested that the PL spectra of titanate nanotubes were also sensitive to the prepared conditions.

As a comparison, the PL spectra excitation at 360 nm is also examined in which $TiO_2$ has normal intrinsic absorption. Figure 5 shows the PL spectra of TANTs heated at different temperatures and P-25. It is found that in this case the PL had a peak at about 420−450 nm with a shoulder peak at about 480 nm, and the PL intensity was lower than that excited by 470 nm. It is obviously seen that the relative intensity of the two peaks was invariant with the increase of temperature. However, the PL intensity of TANTs was irregular with the evolution of temperature. This indicates that their origin was consistent, which results from the intrinsic emission.[30,31] It is also explicit that the PL emission of TANTs excited by 360 nm was completely different from that excited by 470 nm.

In order to further determine the origin of the novel emission, we also checked their photoluminescence excited (PLE) spectra shown in Figure 6 (monitored at 520 nm). From this figure it could be clearly seen that samples a, b, and c (heated at 100, 200, and 300 °C, respectively) had very strong PLE spectra in the visible region (showed clear resonant excitation nature); samples d and e (heated at 400 and 500 °C, respectively) and P-25 had very low PLE intensity in the long-wavelength region and instead had PLE peaks at about 400 nm (interband transition or valence band to conduction band transition). These suggested that the strong emission bands could be directly related to the visible absorption bands of the tubular structure.

By far, we still lacked complete understanding for the origin of the absorption, especially for the three absorption peaks in the visible region. In our former work[26,27] we found that the novel nanomaterials after being vacuum dehydrated had stronger absorption in almost all visible regions and strong PL when excited by visible light. The electron spin resonance (ESR) experiment showed that the materials had a strong ESR signal at $g = 2.0034$, which indicated that special oxygen vacancies existed—so-called "single-electron-trapped" oxygen vacancies (SETOVs). According to those experiments we proposed a possible explanation that just the single-electron-trapped oxygen vacancies formed a sub-band within a forbidden band of titanate nanotubes and resulted in absorption in the visible region.[26,27] The former illumination gave a reasonable explanation for the unfeatured and broad absorption in the visible region. However, illumination could not give a complete understanding for the three absorption peaks in the visible region in the present experiment.

In fact, in our former studies[26,27] we also found that the intensity and band configuration of the absorption spectrum in the visible region markedly changed with the treatment conditions such as the concentration of alkali, acid, pH value, and especially drying conditions and process. We found that the higher the dried temperature was and the longer the time kept, the stronger the intensity of the absorption in the visible region and ESR signal became; moreover, the nanotubes became shorter and shorter, and if the time was enough long, the tubular structure could completely be destroyed and turned to nanoparticles. In order to strictly control the structure and property of the materials, in the present work a special drying process of samples had been performed in which the samples were dried at a low temperature for a long time (60−70 °C for 48 h) before being heated at high temperature for a short time (100−500 °C for 2 h) instead of that dried at high temperature for a long time in our previous experiments. We found that after undergoing the appropriate drying and heating process the three absorption peaks in the visible region could be obtained. We presumed that dehydration of the new nanotubes could take place in different positions (such as the interlayer and intralayer of the nanotubes) in which SETOVs formed had different energy levels and those energy levels could form special bands under a slow and suitable drying and heating process. Further detailed experiments and mechanisms are being done.

## Conclusion

The strong and stable visible absorption bands at 515, 575, and 675 nm and the corresponding photoluminescence of titanic acid nanotubes at room temperature in air were first observed. The special dehydrating process with slow and suitable drying and heating was the key condition that formed the three bands in the visible region. The unusual phenomenon perhaps originated from special oxygen vacancies formed in the dehydration process. These novel nanotubular materials with both strong absorption and photoluminescence in the visible region may be applied to the photocatalysis, nonlinear optics, photoelectric conversion, and devices.

**Acknowledgment.** This work was supported by the Natural Science Foundation of China (No. 90306010), State Key Basic Research "973" Plan of China (No. 2007CB616911), and Program for New Century Excellent Talents in University of China (NCET-04-0653).

**References and Notes**

(1) Fujishima, A.; Honda, K. *Nature* **1972**, *238*, 37.
(2) Khan, S. U. M.; Akikusa, J. *Int. J. Hydrogen Energy* **2002**, *27*, 863.
(3) O'Regan, B.; Grätzel, M. *Nature* **1991**, *353*, 737.
(4) Sauve, G. *J. Phys. Chem. B* **2000**, *104*, 3488.
(5) Nishimura, S. *Appl. Phys. Lett.* **2002**, *81*, 4532.
(6) Castellano, F. N.; Stipkala, J. M.; Friedman, L. A.; Meyer, G. J. *Chem. Mater.* **1994**, *6*, 2123.
(7) Jongh, P. E.; Vanmaekelbergh, D. *Phys. Rev. Lett.* **1996**, *77*, 3427.
(8) Hoffman, M. R.; Martin, S. T.; Choi, W.; Bahnemann, D. W. *Chem. Rev.* **1995**, *95*, 69.
(9) Bach, U.; Lupo, D.; Comte, P.; Moser, J. E.; Weissörtel, F.; Salbeck, J.; Spreitzer, H.; Grätzel, M. *Nature* **1998**, *395*, 583.
(10) McFarland, E. W.; Tang, J. *Nature* **2003**, *421*, 616.
(11) Nishimura, S.; Abrams, N.; Lewis, B. A.; Halaoui, L. I.; Mallouk, T. E.; Benkstein, K. D.; Lagemaat, J.; Frank, A. J. *J. Am. Chem. Soc.* **2003**, *125*, 6306.
(12) Asahi, R.; Morikawa, T.; Ohwaki, T.; Aoki, K.; Taga, Y. *Science* **2001**, *293*, 269.
(13) Khan, S. U. M.; Al-Shahry, M.; Ingler, W. B. *Science* **2002**, *297*, 2243.
(14) Iijima, S. *Nature* **1991**, *354*, 56.
(15) Trentler, T. J.; Hickman, K. M.; Goel, S. C.; Viano, A. M.; Gibbons, P. C.; Buhro, W. E. *Science* **1995**, *270*, 1791.
(16) Shi, W.; Zheng, Y. F.; Wang, N.; Lee, C. S.; Lee, S. T. *Adv. Mater.* **2001**, *13*, 591.
(17) Tang, Z. Y.; Kotov, N. A.; Giersig, M. *Science* **2002**, *297*, 237.
(18) Pan, Z. W.; Dai, Z. R.; Wang, Z. L. *Science* **2001**, *291*, 1947.
(19) Suenaga, K.; Colliex, C.; Demoncy, N.; Loiseau, A.; Pascard, H.; Willaime, F. *Science* **1997**, *278*, 653.
(20) (a) Kasuga, T.; Hiramatsu, M.; Hoson, A.; Sekino, T.; Niihara, K. *Langmuir* **1998**, *14*, 3160. (b) Kasuga, T.; Hiramatsu, M.; Hoson, A.; Sekino, T.; Niihara, K. *Adv. Mater.* **1999**, *11*, 1307.
(21) Zhang, S. L.; Zhou, J. F.; Zhang, Z. J.; Du, Z. L.; Jin, Z. S. *Chin. Sci. Bull.* **2000**, *45*, 1533.
(22) (a) Du, G. H.; Chen, Q.; Che, R. C.; Yuan, Z. Y.; Peng, L. M. *Appl. Phys. Lett.* **2001**, *79*, 3702. (b) Chen, Q.; Du, G. H.; Peng, L. M. *J. Chin. Electron Microsc. Soc.* **2002**, *21*, 265. (c) Zhang, S.; Peng, L. M.; Chen, Q.; Du, G. H.; Dawson, G.; Zhou, W. Z. *Phys. Rev. Lett.* **2003**, *91*, 256103.
(23) Yang, J. J.; Jin, Z. S.; Zhang, J. W.; Zhang, S. L.; Zhang, Z. J. *Dalton Trans.* **2003**, 3898.
(24) Uchida, S.; Chiba, R.; Tomiha, M.; Masaki, N.; Shirai, M. *Electrochemistry* **2002**, *70*, 418.
(25) Adachi, M.; Okada, I.; Ngamsinalapasathian, S.; Murata, Y.; Yoshikawa, S. *Electrochemistry* **2002**, *70*, 449.
(26) Qian, L.; Jin, Z. S.; Zhang, J. W.; Huang, Y. B.; Zhang, Z. J.; Du, Z. L. *Appl. Phys. A: Mater.* **2005**, *80*, 1801.
(27) Zhang, S. L.; Li, W.; Jin, Z. S.; Zhang, Z. J.; Du, Z. L. *J. Solid State Chem.* **2004**, *177*, 1365.
(28) Wendlandt, W. W.; Hecht, H. G. *Reflectance Spectroscopy*; Wiley-Interscience: New York, 1966.
(29) Anderson, J. R.; Pratt, K. C. *Introduction to Characterization and Testing of Catalysts*; Academic Press: Australia, 1985.
(30) Tang, H.; Berger, H.; Schmid, P. E.; Levy, F. *Solid State Commun.* **1993**, *87*, 847.
(31) Saraf, L. V.; Kshirsager, S. T. *Int. J. Mod. Phys. B* **1998**, *12*, 2635.